--------------------------------------------------------------------------------

\documentstyle [12pt,a4]{article}

\newcommand{\be}{\begin{equation}}
\newcommand{\ee}{\end{equation}}
\newcommand{\bea}{\begin{eqnarray}}
\newcommand{\ena}{\end{eqnarray}}
\newcommand{\sect}[1]{\setcounter{equation}{0}\section{#1}}
\newcommand{\vs}[1]{\rule[- #1 mm]{0mm}{#1 mm}}

\newcommand{\sm}[2]{\frac{\mbox{\footnotesize #1}\vs{-2}}
                   {\vs{-2}\mbox{\footnotesize #2}}}

\newcommand{\shalf}{\sm{1}{2}}

\newcommand{\cd}{\cal{D}}

\newcommand{\prt}{\partial}

\newcommand{\al}{\alpha}
\newcommand{\var}{\varphi}

\newcommand{\NP}[1]{Nucl.\ Phys.\ {\bf #1}}
\newcommand{\PL}[1]{Phys.\ Lett.\ {\bf #1}}

\newcommand{\PR}[1]{Phys.\ Rev.\ {\bf #1}}
\newcommand{\PRL}[1]{Phys.\ Rev.\ Lett.\ {\bf #1}}
\newcommand{\MPL}[1]{Mod.\ Phys.\ Lett.\ {\bf #1}}

\newcommand{\JMP}[1]{J.\ Math.\ Phys.\ {\bf #1}}

\begin{document}
\renewcommand{\thefootnote}{\fnsymbol{footnote}}
\newpage
\setcounter{page}{0}

\vs{20}

\begin{center}

{\Large {\bf 3D-Ising Model as a String Theory \\[.5cm]
in Three Dimensional Euclidean Space}}\\[2cm]

{\large A. Sedrakyan\footnote{permanent address : Yerevan Physics Institute,
Br. Alikhanian, st.2, Yerevan 36, Armenia}}\\
{\em{Laboratoire de Physique Th\'eorique}}
{\small E}N{\large S}{\Large L}{\large A}P{\small P}
\footnote{URA 14-36 du CNRS, associ\'ee \`a l'E.N.S. de Lyon, et au L.A.P.P.
(IN2P3-CNRS) d'Annecy-le-Vieux}
\\
{\em Chemin de Bellevue, BP 110, F - 74941 Annecy-le-Vieux Cedex,
France}\\[.5cm]

\end{center}
\vs{20}

\centerline{\bf Abstract}

\indent

A three dimensional string model is analyzed in the strong coupling regime.
The contribution of surfaces with different topology to the partition function
is essential. A set of corresponding models is discovered. Their critical
indices,which depend on two integers $(m,n)$, are calculated analytically. The
critical indices of the three dimensional Ising model should belong to this
set. A possible connection with the chain of three dimensional lattice Pott's
models is pointed out.

\vs{20}

\rightline{{\small E}N{\large S}{\Large L}{\large A}P{\small P}-A-410/92}
\rightline{November 1992}

\newpage

\renewcommand{\thefootnote}{\arabic{footnote}}
\setcounter{footnote}{0}
\newpage

\noindent

\sect{Introduction}

\indent

The understanding and construction of strings in noncritical dimensions is
presently attracting interest of string theorists. For example, the string
theory in four dimensional space-time is probably a good framework for a theory
of strong interactions of the elementary particles. This framework first
appeared at the end of seventies,motivated by the dual resonance approach
\cite{1}. Moreover, the three dimensional Ising model (3DIM), being a gauge
theory with the group $Z_2$, is, in fact, a theory of random surfaces. Polyakov
has put forward the idea \cite{2} of its equivalence with a fermionic string
near the point of the second order phase transition. An important element in
understanding of the 3DIM as a string theory is the fact that closed surfaces
with different topologies contribute to the partition function, and that the
module of the string coupling constant is one. Thus, we have a string in the
strong coupling regime.

There have been many attempts to construct a fermionic string on the lattice
\cite{2}-\cite{9} which corresponds to the 3DIM, but the construction
suggested in \cite{8} differs essentially from  the others. There, a naive
continuum limit (lattice spacing $a \rightarrow 0$) of the lattice action
exists and three dimentional Dirac fermions appear in the action quadratically.
 This fact allowed the author to develop in \cite{10} the idea of equivalence
of 3DIM with the theory of some kind of matter fields which interact with 2D
quantum gravity.

The essential ingredient of the approach considered in \cite{10} is the
evaluation of the contribution of the two dimensional manifolds, immersed into
3D-euclidean space and singular at the end points of selfintersection lines,
to the functional integral over all surfaces (partition function of the 3DIM).
The singular surfaces are essential for the 3DIM, because some of them appear
in the partition function with the weight -1, ensuring cancellation of the
contribution of a part  of surfaces. On the other hand, some of singular
surfaces are nothing else but unoriented surfaces immersed into 3D-euclidean
space (sphere with  M\"obius cups).

It happens that the spins of Dirac fermions in the presence of singularities
are modified and, also, that the original vacuum must be changed by filling it
with them. The change of the spin of the fermions diminishes the central charge
 of the matter fields and that is why the KPZ equation \cite{11}  for the
$SL(2.R)$ coupling constant of the 2D quantum gravity, which is the condition
of restoration of the reparametrization invariance has a solution for $D=3$.

In this article I shall further develop the approach presented in ref \cite{10}
correcting at the same time the errors made in that article.  It will be shown
that the matter sector of the model, consists of three scalars (surface
coordinates $\vec{X}$) and a spin (0,1) fermionic $(b,c)$ system (corresponding
to singularities) with their total central charge equal to one. In conformal
gauge the 2D-gravity represented by Liouville acton and the ghost fields has to
be introduced. The sum over all possible singular surfaces (and at the same
time over all topologies) in the partition function induces a new term in
effective action of the model while keeping the vacuum unchanged, namely,
inteaction of Liouville field with the spin (0,1) bosonized fermionic $(b,c)$
system. The true vacuum now differs from the original one which corresponded
to a simple topology. The set of models is discovered, for which the
contribution of the singular surfaces in partition function is essential. Then
the scaling behaviour will be analyzed and the critical indices, which depend
on two integers $(m,n)$, analitically calculated. The critical indices of the
3DIM should belong to this set.

It is known \cite{11},\cite{22}, that $c=1$ string theory lies on the boundary
between the "weak" and "strong" interacting phases of 2D quantum gravity.
Perhaps the approach presented here may open a way to cross the $c = 1$
barrier.

\sect{2D-gravity structure of the 3DIM}

\indent

In order to reveal the structure of 3DIM near the critical point we shall work
with the lattice formulation of the fermionic string, where the three
dimensional Dirac fermions $\psi_L$ and $\psi_R$ are placed in the middle
points
 of links of the  lattice as explained in ref \cite{8}. The general ideology
is that of ref. \cite{10}.

As it was shown in ref. \cite{8}, this action reproduces exactly the 3DIM
partition function at arbitrary temperature, along with the correct sign-factor
for the surfaces. The classical continuum limit (lattice spacing $a \rightarrow
 0$) is
\be
S_0 = \lambda \cdot arrea + \frac{i}{2} \int d^2 \xi \sqrt{g} \bar{\psi} (\xi)
(\gamma^\al \vec{\prt}_\al - \vec{\prt}_\al \gamma^\al) \psi(\xi) .
\label{eq:2.1}
\ee
In eq.(\ref{eq:2.1}) $\bar{\psi}, \psi$ are 3D Dirac fields living on the
two-dimensional surface $\vec{X}(\xi)$ ; $g_{\al \beta} = \prt_\al \vec{X}
\cdot \prt_\beta \vec{X}$  is an induced metric , $\gamma_\al = \prt_\al
\vec{X} \cdot \vec{\sigma}$ ($\vec{\sigma}$ are the Pauli matrices) and
$\lambda =\frac{1}{a^2} \ln \ th \ J/KT$, where $J$ is the 3DIM coupling
constant and $T$ is the temperature.

Here, the following remark is in order. In ref. \cite{8}, taking classical
continuum limit of the lattice action, we did not distinguish left (or right)
fermions, placed at the opposite links of the plaquettes (see fig.3 in
\cite{8}) and therefore translational invariance of the theory was broken. This
 distinction should be made, as one can see by making diagonalization of the
action in the momentum space of a flat manifold. However, we will then have a
complicated expression for the action in the continuum limit even in a flat
case, which is free of anomalies, but like the Nambu action has some other
problems. By identifying some of the fields we obtain a simple massless action
(\ref{eq:2.1}) which now has gravitational (or $SO(3)$ rotational) anomalies,
absent in 3DIM. Following the general procedure we can correct the situation
by adding the 2D-gravity action induced (due to anomalies) by quantum
fluctuations of $\psi$ (and also of $\vec{X}$) and restore the original
symmetries of the theory \footnote{I thank A. Polyakov for a criticism at this
point.}. In principle, one is doing the same, when replacing the Nambu action
for bosonic string by the Polyakov's action.

Therefore, the candidate for the action of 3DIM near the critical point is
\be
S_1=\lambda \cdot area + \frac{i}{2} \int d^2 \xi \sqrt{g} \bar{\psi}
\Omega^{-1} \sigma^a e^\al_a (\prt_\al + \Gamma_\al) \Omega \psi + W_{grav}
(\vec{X}) + S(ghost)
\label{eq:2.2}
\ee

In (\ref{eq:2.2}) we have made an algebraic transformation of the fermionic
action (\ref{eq:2.1}) (see \cite{12} for details).Here, ${e_a}^\al$ and
$\Gamma_\al$ are the zweibeins and $SO(2)$ spinor connection, corresponding
to induced metric, $\Omega$ is the element of Spin(3) and defines the matrix
which rotates ortonormalized Frene basic vectors of the surface into the flat
ones.The 2D-gravity action $W_{grav}(\vec{X}$) with the $SL(2,R)$ coupling
constant $k$ (or with the charge $Q_L$ in conformal gauge) should be added,
ensuring restoration of all anomalous symmetries of the theory. This
corresponds to the KPZ \cite{11} equation $C_{tot}=0$.

Now we should investigate the contribution of Whitney singularities of the two
dimensional surfaces, immersed in 3D euclidean space in the string functional
integral. Let us represent $\Omega$ as
\be
\Omega = \Omega_s \cdot \Omega_r  ,
\label{eq:2.3}
\ee
where $\Omega_r$ is a regular rotation matrix, making a surface flat, but not
affecting singularities and $\Omega_s$ is the singular part of $\Omega$.

Since the rotation and reparametrization invariances are restored by addition
of $W_{grav}$, we can represent the fermionic action in the form ($\Omega_r$
dissapears and metric becomes flat)
\be
\frac{i}{2} \int d^2 \xi \bar{\psi} \Omega^{-1}_s \sigma^a \prt_a \Omega_s
\psi     .
\label{eq:2.4}
\ee
It is easy to calculate $\Omega_s$. In the case of flat surface the Whitney
singularity looks like in Fig.1.

\vs{70}

\begin{center}

Fig. 1

\end{center}

One can parametrize the surface as follows
\[
\begin{array}{ll}
X_3=0,& w=z^2, \nonumber\\
w=X_1 + iX_2, & z= \xi_1 + i\xi_2 .
\nonumber
\end{array}
\]

The straightforward calculations, by use of formula (2.7) in
\cite{10}, gives
\bea
\Omega^{-1}_s \prt \Omega_s &=& \frac{1}{4z} \sigma_3 \nonumber \\
\Omega^{-1}_s \bar{\prt} \Omega_s &=& -\frac{1}{4\bar{z}} \sigma_3
\ena
and
\be
\Omega_s = \left( \frac{z}{|z|} \right)^{-1/2} ,
\label{eq:2.6}
\ee
which means that there is a magnetic flux in the point.

Then, redefining the fields
\be
b= \bar{\psi_L} \Omega^{-1}_s , c = \Omega_s \psi_L
\label{eq:2.7}
\ee
one obtains a free action
\be
\frac{1}{\pi} \int d^2 z b \prt c + c.c.
\label{eq:2.8}
\ee
The spins of b, c fields are equal to $\frac{1}{2} (1 \pm Q)$, where
$Q=1$.\footnote{Here, I would like to mention that in ref. \cite{10} a mistake
of taking $Q=k/2$ has been made. Along with a renormalization of the coupling
constant $k$ the field $h$ has also to be renormalized. This should be done
consistently with the fact, that the topological origin of the sign-factor of
3DIM (which takes values $\pm 1$) prohibits its renormalization. This imposes
the condition $\sqrt{\frac{k}{2}}dh^{ren} \simeq \delta^{(2)} (\xi)$, which
implies that $Q=1$, instead of $Q=k/2$. Besides, there is no need to introduce
an additional fermion.}

Thus, in the presence of vortices (singularities) the conformal spin of
fermions changes and the correspodning central charge of the conformal theory
becomes $1-3Q^2$.

Besides modifying the spin of the fermions, as was shown in \cite{10}, one
should fill the vacuum state of the system with the $(b,c)$ pairs and the
corresponding correlator will appear in the expression for free energy.

This means that when acting by $(b,c)$  operators on the vacuum state we
create the singularities on the sphere transforming it into the surface with a
different topology.

Next, following ref. \cite{10}, instead of induced metric $g_{\al \beta} =
\prt_\al \vec{X} \prt_\beta \vec{X}$, we introduce the independent zweibeins
${e_a}^\al$ and fix the gauge. The ghost fields will appear. For our purpose,
to calculate the critical indices of the system, the conformal gauge $g_{\al
\beta} = e^\var \delta_{\al \beta}$ and the technic, developed by F. David
\cite{14} and J. Distler, H. Kawai \cite{15} (DDK), seem to be most
appropriate.

Finally, the free energy $F$ of the 3DIM near the critical point is
\bea
F &=& \sum^\infty_{N=0} \int {\cd} \vec{X} {\cd} (b,c) {\cd} ghost {\cd} \var
\cdot \frac{1}{(N!)^2} \cdot \nonumber \\
&& \left( \int d^2z b(z) \int d^2 z' c(z') \right)^N e^{-S}    , where
\label{eq:2.10}
\ena
\[
S =  \frac{1}{2 \pi} \int d^2z \prt \vec{X} \bar{\prt} \vec{X} +
\frac{1}{\pi} \int d^2z  b \bar{\prt} c + c.c. + S(ghost)
\]
\be
+ \frac{1}{2\pi} \int d^2z  ( \prt \var \bar{\prt} \var - \frac{1}{4}
Q_L \sqrt{g} \hat{R} \var ) + \mu \int d^2z  e^{A\var} \label{eq:2.11}
\ee
\[
\mu = \lambda_0 - \lambda, \ \ \ \ A = - \frac{Q_L}{2} +
\sqrt{Q_L^2 -8},
\nonumber
\]

In eq.(\ref{eq:2.10}), $N$ is the number of pairs of singularities of the
surface, the presence of the factor $(N!)^2$ is a consequence of having $N$
identical $b$ and $N$ identical $c$ fields, $\lambda_0$ is bare cosmological
constant and $Q_L$ is the Liouville background charge. The expression for
$Q_L$, in terms of $SL(2,R)$ couling cosntant $k$, is
\be
{Q_L}^2 = \frac{k}{k+2} - 2k -1  .
\label{eq:2.12}
\ee

The KPZ equation \cite{11}
\be
c_{tot} = 3+(1-3Q^2) - 26 + (1+3Q^2_L)=0   ,
\label{eq:2.13}
\ee
which is the condition for restoration of all anomalous symmetries, can be
solved. For $Q=1$ we obtain $k=-3$ and $Q_L = 2 \sqrt{2}$. The corresponding
central
charge of the matter fields of the theory is equal to one.

\sect{Scaling properties. Specific heat index $\al$}

\indent

In order to calculate the critical indices consider the bosonized anticommuting
$(b,c)$ system \cite{16}. The action is
\be
\frac{1}{4} \int d^2z  b \bar{\prt} c+c.c. = \frac{1}{2 \pi} \int d^2z  (\prt
\phi \bar{\prt} \phi - \frac{i}{4} Q \hat{R} \phi)
\label{eq:3.1}
\ee
where $b=e^{-i\phi(z)}$ and $c=e^{i \phi (z)}$.

As they appear in vacuum (see (\ref{eq:2.10})), the $(b,c)$ fields are
polarized by the gravitational filed $\var$ and also by $\phi$
(selfpolarization),so that the ground state is reparametrization invariant.
After this dressing,
\be
b=e^{B\var} e^{iD\phi}, \ \ \ c=e^{\bar{B} \var} e^{i \bar{D} \phi},
\label{eq:3.2}
\ee
the conformal dimensions of both $b$ and $c$ field should be (1,1):
\bea
- \frac{B(B+Q_L)}{2} &+&  \frac{D(D+Q)}{2} =1 \nonumber \\
- \frac{\bar{B}(\bar{B}+Q_L)}{2} &+&  \frac{\bar{D}(\bar{D}+Q)}{2} =1
\label{eq:3.3}
\ena

Let us introduce the string coupling constant $\Lambda$, by writing
$\Lambda^{2N}$ in front of the exponent in (\ref{eq:2.10}) and take the sum
in $F$ over $(b,c)$ pairs. In order to simplify the expression for the
effective action $S_{eff}$ we change the sum over $N$ of the series in
(\ref{eq:2.10}) into the
product of two independent exponential series for $b$ and $c$. One can do that,
because the condition of mutual cancellation of the $(b,c)$, cosmological and
background charges (see (\ref{eq:3.5})) will select the equal number of $b$ and
$c$ fields in the expression for the path integral (\ref{eq:2.10}). Then, from
eq. (\ref{eq:2.10}), we obtain:
\be
S_{eff} = S + \Lambda \int d^2z e^{B \var} e^{iD \phi} + \Lambda \int d^2z
e^{\bar{B}\var} e^{i \bar{D} \phi}
\label{eq:3.4}
\ee
This action describes a Liouville theory interacting with a bosonized fermionic
$(b,c)$ system. What is the geometrical meaning of the presence of $b$ and $c$
fields in the vacuum ? Appearence of one pair of $(b,c)$ in front of exp in
(\ref{eq:2.10}) is equivalent to a change of vacuum charges $Q(1-g)$ and
$Q_L (1-g)$ ($g$ is the genus of Riemann surface) of the fields $\phi$ and
$\var$ to $Q(1-g+D+\bar{D})$ and $Q_L(1-g+B+\bar{B})$ respectively, and can be
interpreted as a change of $g$.

There are two types of singularities of Riemann surfaces in 3D-space. One
changes the topology (Fig.2a) and the other (Fig. 2b) does not.

\vs{70}

\begin{center}
Fig.2

\end{center}

In 3DIM the surfaces with all topologies are present in the partition function.
 Therefore, the expression (\ref{eq:3.4}) should contain the dressed $(b,c)$
system, whose
appeareance in the vacuum is equivalent to a change of genus by $\shalf$ (as in
Fig.2a).

Let us take $n$ pairs of $(b,c)$ fields, which, together with $m$ cosmological
terms, create the M\"obius cap on the manifold, as in Fig. 2a, which
corresponds to the following equations
\bea
&& n(D+\bar{D}) = -Q/2 \nonumber \\
&& n(B+\bar{B}) + mA = - Q_L/2           .
\label{eq:3.5}
\ena

Equations (\ref{eq:3.3}) and (\ref{eq:3.5}) determine $B, \bar{B}, D, \bar{D}$
as functions of integers $m$ and $n$.

Following DDK \cite{14,15}, we can evaluate specific heat index $\al$ by a
simple scaling argument. Consider the scaling transformations
\be
\phi \rightarrow \phi+ \phi_0, \ \  \var \rightarrow \var + \frac{\var_0}{A},
 \ \ \mu \rightarrow \mu e^{-\var_0} \ \ \mbox{ and } \ \ \Lambda \rightarrow
\Lambda
e^{-x}
\label{eq:3.6}
\ee
with constant $\phi_0, \var_0$. In order to find the scaling behaviour we
should demand the invariance of the interacting terms in eq. (\ref{eq:3.4}),
which implies
\bea
&& - x + \frac{B}{A} \var_0 + iD \phi_0 =0 \nonumber \\
&& - x + \frac{\bar{B}}{A} \var_0 + i\bar{D} \phi_0 =0   .
\label{eq:3.7}
\ena
Solving these equations, along with (\ref{eq:3.3}) and (\ref{eq:3.5}), one
finds
\be
i \phi_0/\var_0 = \frac{Q}{A^2(-\frac{Q_L}{A} + \frac{m}{n-1/2})} =
\frac{1}{2(2+\frac{m}{n-1/2})}
\label{eq:3.8}
\ee
and
\be
\frac{x}{\var_0} = p=\frac{1}{2n} \left( 1-m- \frac{1}{4(2+\frac{m}{n-1/2})}
\right)   .
\label{eq:3.9}
\ee

Then, using the scaling transformations, it is not hard to see that
\be
S_{eff} \rightarrow S_{eff} - \frac{Q}{A} \var_0 - i Q \phi_0  .
\label{eq:3.10}
\ee
Equivalently
\be
F(\mu e^{- \var_0}, \Lambda e^{-x}) = e^{(\frac{Q}{A} + Q
\frac{i\phi_0}{\var_0}) \var_0} F(\mu,\Lambda)     ,
\label{eq:3.11}
\ee
i.e.
\be
F(\mu ,\Lambda) = \mu^{-(\frac{Q}{A} + Q
\frac{i\phi_0}{\var_0})} F \left( 1,\frac{\Lambda}{\mu^{x/\var^0}} \right)
= \mu^{2-\al} \tilde{F} \left( \frac{\Lambda}{\mu^p} \right)   .
\label{eq:3.12}
\ee

Thus, the specific heat (or string susceptibility) index is (use also eq.
(\ref{eq:3.8}))\footnote{For a review of the definition and old numerical
calculations see, for example, \cite{17}.}:
\be
\al = \frac{1}{2(2+ \frac{m}{n-1/2})}
\label{eq:3.13}
\ee
and $\Lambda \sim \mu^p$. Since $m \geq 1$ implies $p<0$, we are in the strong
coupling regime.

If we introduce dressed $(b,c)$ operators, which do not create topology (case
of Fig.2b) and which depend on integers $(m',n')$, their scaling equations
(eq. like (\ref{eq:3.7})) should be consistent with eq. (\ref{eq:3.7}). It is
easy to find that the condition of consistency is $\frac{m'}{n'} =
\frac{m}{n-1/2}$ and to see that the specific heat index $\al$ does not change.

\sect{Correlation length index $\nu$}

\indent

In order to calculate $\nu$ one should investigate long distance
behaviour of the correlator of some operators $\Phi (\vec{X})$, placed at
distance $\vec{X}$. Then the correlation length and its index $\nu$ are defined
 as follows
\be
\frac{\int d \vec{X} (\vec{X})^2 <\Phi (\vec{X}) \Phi (0) >}
{\int d \vec{X} <\Phi (\vec{X}) \Phi (0) >} \equiv \xi^2 \sim \mu^{-2 \nu}  .
\label{eq:4.1}
\ee

This expression shows that we need to find the anomalous scaling dimension of
$\vec{X}$. In the string picture the correlation length of operators in an
external space is calculated by insertion of the operators of the type
\be
\Phi (\vec{X},\cdot) =  \int d^2 \xi \Phi (\cdot) \delta^{(D)} (\vec{X} -
\vec{X} (\xi))
\label{eq:4.2}
\ee
into the partition function integral.

To find the anomalous scaling dimension of $\vec{X}$ and then to calculate the
correlation length index $\nu$ it is enough to consider the world-sheet
operator {\bf 1}, with the conformal dimension 0. It seems to us, that in the
presence of background Liouville and $(b,c)$ charges $Q_L$ and $Q$ that is
\be
{\bf 1} = e^{-Q_L \var} e^{-iQ\phi}
\label{eq:4.3}
\ee

Imposing the scale invariance of the
\be
e^{-Q_L \var} e^{-iQ\phi} \cdot \delta^{(3)} (\vec{X} - \vec{X} (\xi))
\label{eq:4.4}
\ee
under the transformations (\ref{eq:3.6}) along with $\vec{X} \rightarrow \rho
\vec{X}$ and using the scaling arguments employed in the previous section we
obtain:
\be
\nu = \frac{2-\al}{3}.
\label{eq:4.5}
\ee

\sect{Discussion of results}

\indent

We have obtained sets of critical indices which depend on two integers, $m$ and
$n$. The common origin of all of them is the contribution of singular surfaces
to the partition function  of random surfaces. Among them one can find the
critical indices of 3DIM.

An enormous and diverse work has been done on numerical calculations of 3DIM
critical indices. A summary  can be found in an article by J.-C. Le Guillou
and J. Zinn-Justin \cite{18}. The results quoted in this reference are:
\be
\al=0.11, \ \ \gamma=1.239, \ \ \nu = 0.631, \ \ \beta=0.327, \ \ \eta=0.0375
\label{eq:5.1}
\ee

We can pick a choice of integers $m$ and $n$, for example $m=4$, $n=2$, for
which $\al \sim 0.107, \nu \sim 0.631$ and obtain the results within
experimental errors. However, it seems to us that the case $m=1, n=1$, for
which
\be
\al=1/8, \ \ \nu =0.625
\label{eq:5.2}
\ee
is more natural. These values coincide with some old estimates \cite{17}. They
also agree with a recent Monte Carlo renormalization group study of
indices \cite{19}, where $\nu =0.624(2)$ and $\eta = 0.026(3)$ (we see, that
the hardest index $\eta$ becomes essentially smaller than one in
(\ref{eq:5.1})).

The case $m=0, n=1$ with the indices
\be
\al = 1/4, \nu = 7/12
\label{eq:5.3}
\ee
is in good agreement with the three dimensional indices for a self-avoiding
walk(SAW) problem \cite{20}. In the limit, where $n=1, m \rightarrow \infty$
the
indices get the values
\be
\al =0 \ \ \mbox{ and } \ \ \nu =2/3,
\label{eq:5.4}
\ee
which are in a good agreement with the 3D $U(1)$-model \cite{13,21}. One can
conjecture that the series $n=1$ and arbitrary $m$ corresponds to 3D Pott's
models.

At the end of this analysis I would like to add a few words about magnetic
susceptibility index $\gamma$. In order to calculate the magnetic
susceptibility $\gamma$ we need to analize the response of the system to the
presence of the magnetic field $h$, i.e. to the presence of the additional
term, $h\sum_i \sigma_i$, in the original lattice action of the 3DIM. Then
$\gamma$ is defined by the anomalous scaling dimension of spin field $\sigma$.
It is possible to develope a rough arguments (see also \cite{17}) which gives
the result $\gamma =2 \nu$ (equivalently $\eta =0$), but, to obtain a  precise
answer, one should represent the original spin variable $\sigma$ in terms of
stringy fields and calculate complicated correlators, as suggested in ref.
\cite{2}. In principle, the direct and more precise numerical calculation of
$\eta$ can be decisive for final determination of the critical indices.

Here we have calculated the indices starting from the spherical topology, but
all nonorientable surfaces are taken into account by use of singularities. As
for orientable ones, the carefull analysis of operator algebra is needed. The
first impression is that the consistent introduction of operators which
increase the genus for one, does not change the scaling behaviour.

Presumably, like two dimensional minimal models interacting with gravity, this
set of models also corresponds to some topological field theory. This is an
interesting question and may be relevant to finding a correct topological
definition of the sign-factor of the 3DIM.

In conclusion, I would like to make the last remark. It seems to me that the
picture presented here can be interpreted in the following way. We have a
matter field $\vec{X}$ interacting with the ordinary Liouville field
(representing the fluctuations of metric of 2D gravity) and also interacting
with another "Liouville" field originating from the bosonized fermionic
$(b,c)$ system and representing the "fluctuations" of topology. (There is also
an interaction between two Liouville fields). It is tempting to conjecture that
this interpretation of these two Liouville fields may be valid for a general
noncritical strings.

I would like to thank for  discussions J.~Ambjorn, E.~Buturovi\'c, D.~Gross,
A.~Kavalov, I.~Klebanov, Y.~Kogan, V.~Kazakov, J.-C. Le Guillou, M.~Mkrtchian,
H.~Verlinde, E.~Verlinde and the theory division of LAPP, where this work was
finished. I especially acknowledge A.~Polyakov for  discussions during many
years and for criticism. I would especially like to thank J.~Distler,
conversations with whom stimulated my interest to the DDK approach to
2D-gravity.

\end{document}